%% file: PRD-Version-3.tex
\documentclass[preprint,aps,prd,showpacs,nofootinbib,superscriptaddress,floatfix,showkeys]{revtex4-2}
\usepackage{amsfonts}
\usepackage{amsmath}
\usepackage{amssymb} 
\usepackage{bm} 
\usepackage{color}
\usepackage{enumerate}
\usepackage{graphicx}  
\usepackage{dcolumn}
\usepackage{hyperref}
\usepackage{lineno}

\hypersetup{
	colorlinks=true,       
	linkcolor=blue,          
	citecolor=magenta,        
	filecolor=cyan,      
	urlcolor=magenta           
}

\labelformat{section}{Section #1} 
\labelformat{subsection}{Section #1} 
\labelformat{figure}{Fig.~#1} 
\labelformat{subfigure}{Fig.~\thefigure#1} 
\labelformat{table}{Tab.~#1} 
\begin{document}
\title{Maximal chirality transfer in the photon-graviton conversion in the early universe} 
\author{Ashu Kushwaha} 
\email{ashuk@iisc.ac.in}
\author{Rajeev Kumar Jain}
\email{rkjain@iisc.ac.in}
\affiliation{Department of Physics, Indian Institute of Science C. V. Raman Road, Bangalore 560012, India}
%
	%
	%
\begin{abstract}

While photons and gravitons do not interact significantly, photons can be converted to gravitons in a background magnetic field -- a phenomenon known as the Gertsenshtein effect. 
In this paper, we investigate whether chiral electromagnetic (EM)  waves can be converted to chiral gravitational waves (GW) in the presence of primordial magnetic fields during the radiation-dominated epoch of the early universe. We consider two situations wherein chirality is either present in the propagating EM waves or it exists in the background magnetic field. 
Our analysis shows that while the conversion probability increases with stronger magnetic fields, it remains insensitive to the chiral nature of the background magnetic field. Consequently, the net chirality parameter is independent of the chirality of the background field in both cases.
Finally, we demonstrate that the present-day energy density of the produced chiral GWs peaks at a frequency of $\sim 100$ GHz, and the corresponding characteristic strain can be sensitive to current and future missions designed to detect high-frequency GWs.

\end{abstract}
\maketitle
\newpage

\section{Introduction}
Gravitational waves (GW) and electromagnetic (EM) waves are two powerful and independent tools for probing the physics of the very early Universe. Primordial GWs and the cosmic microwave background (CMB) as relic EM radiation offer profound insights into the history of cosmic evolution.
In a remarkable paper~\cite{1962-Gertsenshtein-JETP}, Gertsenshtein showed that the EM waves propagating through a transverse background magnetic field in curved space could be converted into  GWs -- an effect known as the \emph{Gertsenshtein effect}. 
It was soon recognized that the inverse process is also possible with the conversion occurring in an oscillatory manner~\cite{Boccaletti1970,1974-Zeldovich-SJETP}. Although this effect is entirely classical, it also can be understood within the framework of quantum mechanics, analogous to neutrino flavor mixing (or axion-photon conversion~\cite{1988-Raffelt.Stodolsky-PRD,2001-Deffayet.etal.Zaldarriaga-PRD}), wherein the background magnetic field (spin-1) facilitates resonant mixing between a photon (spin-1) and a graviton (spin-2)~\cite{2023-Palessandro.Rothman-PDU}. The background field provides the necessary angular momentum required for the mixing and also determines the efficiency of the conversion. 
Since gravity (or GWs) interacts very weakly with radiation (EM waves), the background magnetic field must be sufficiently strong to achieve the observable effects. 
Such strongly magnetized regions exist across various length scales in the Universe, such as in neutron stars, white dwarfs, magnetars, and the early Universe. 
These regions have served as efficient natural laboratories for exploring the phenomenological consequences of the conversion mechanisms, for instance, see Refs.~\cite{2018-Hook.etal-PRL,2021-Witte.etal-PRD} for the axion-photon conversion in neutron stars. The photon-graviton conversion, in particular, has been used to constrain properties of primordial magnetic field (PMF) by analyzing the conversion of CMB photons into gravitons~\cite{Chen:1994ch,Cillis:1996qy,2013-Chen.Suyama-PRD,2013-Ejlli-PRD}. In Refs.~\cite{Bastianelli:2004zp,Bastianelli:2007jv}, authors have also studied the photon-graviton conversion at the one-loop level.
	
Various observational probes have confirmed the presence of magnetic fields in the galaxies and galaxy clusters with the typical strength of ${\cal O} (\mu$G) coherent over the scales of kpc to Mpc~\cite{1994-Kronberg-Rept.Prog.Phys.,1996-Beck.etal-ARAA,2001-Grasso.etal-PhyRep,2002-Widrow-Rev.Mod.Phys.,2004-Vallee-NAR,2004-Gaensler.Beck.etal-AstroRev,2004-Giovannini-IJMPD,2007-Barrow.etal-PhyRep,2011-Kandus.Kunze.Tsagas-PhysRept,2013-Durrer.Neronov-Arxiv,2016-Subramanian-Arxiv}. The origin of these magnetic fields is still unknown.
Moreover, the FERMI and High Energy Stereoscopic System (HESS) measurements of GeV $\gamma$-rays from TeV blazars provide a lower bound of $\sim 10^{-16}\,\rm{G}$ on the magnetic fields in intergalactic voids with a coherence length of Mpc or larger~\cite{2010-Neronov.Vovk-Sci}. This suggests that these magnetic fields might be of primordial origin, which are generated in the early universe, such as during inflation and reheating~\cite{1988-Turner.Widrow-PRD,1991-Ratra-Apj.Lett,2007-Martin.Yokoyama-JCAP,2011-Byrnes.Jain-JCAP,Ferreira:2013sqa,Ferreira:2014hma,Kobayashi:2014sga,Fujita:2019pmi,2019-Kushwaha.Shankaranarayanan-PRD,2020-Bamba.Odintsov.etal-JCAP,2021-Giovannini-JCAP,Tripathy:2021sfb,Tripathy:2022iev}, phase-transition~\cite{1991-Vachaspati-PLB,1993-Dolgov-PRD,Durrer:2003ja}, etc, and later amplified due to dynamo mechanism. Amongst these scenarios, the inflationary mechanisms generate sufficiently strong magnetic fields with large coherence lengths, which could even be larger than the horizon size. As we shall discuss later, a necessary requirement for the photon-graviton conversion to happen efficiently is that the photon wavelength must be smaller than the coherence length of the background magnetic field. The EM field (massless photon) has two transverse degrees of freedom --- left- and right-circular polarizations, which are associated with the left- and right-handed helicity modes. When both polarization modes evolve identically (i.e., following the same dispersion relation), the resulting EM field is non-helical (or non-chiral). In contrast, if the modes propagate differently, such as one being amplified while the other is attenuated, the EM field becomes helical (or chiral), resulting in a net non-zero helicity~\cite{2011-Durrer.Hollenstein.Jain-JCAP,2020-Kushwaha.Shankaranarayanan-PRD}. The presence of magnetic fields in the early universe influences various cosmological processes, providing an avenue to constrain their properties~\cite{2019-Jedamzik.Saveliev-PRL,2021-Kushwaha.Shankaranarayanan-PRD}.
	
In the era of precision cosmology, the primary focus of various ongoing/future experiments is to probe the fundamental physics at different scales. The direct detection of GWs by the LIGO and Virgo collaborations~\cite{GW150914-LIGOScientific}, has provided us with a very powerful tool to understand the nature of compact objects. Over the last decades, many GW missions are proposed in different broad frequency ranges, e.g., ground-based interferometers like KAGRA~\cite{Somiya:2011np-KAGRA} have sensitivity in the frequency range $1-1000 \, {\rm Hz}$, the pulsar timing arrays such as EPTA~\cite{Kramer:2013kea-EPTA} and NANOGrav~\cite{McLaughlin:2013ira} can observe GWs at nano-Hertz frequency range. The space-based interferometers like LISA~\cite{Amaro-Seoane:2012aqc,Bartolo:2016ami-LISA}, DECIGO~\cite{Kawamura:2011zz-DECIGO} are sensitive at ${\rm mHz}- {\rm Hz}$. Several physical systems can produce GWs at very broad and different frequency ranges,~$10^{-18}-10^{10}$~\cite{1999-Maggiore-PhyRept,2004-Bisnovatyi.Kogan-CQG,2006-Cruise-CQG,2012-Cruise-CQG,2009-Sathyaprakash.Schutz-LivRevRel}, and direct measurements of the GWs can directly probe the underlying physics of these sources. 
So far, all these missions are dedicated to low-frequency ranges. Due to technological limitations and challenges, the direct detection techniques for GWs in high-frequency ranges (MHz to a few GHz) are still at a preliminary stage. Recently, some high-frequency gravitational wave (HFGW) detectors have also been proposed, and a few of them are operational, see Refs.~\cite{2017-Chou.etal-PRD,2008-Nishizawa.etal-PRD,2020-Aggarwal.etal-LivingRevRel,Domcke:2022rgu,Bringmann:2023gba}. These detectors could play a crucial role in exploring physics beyond the Standard Model, such as primordial black holes (PBHs), exotic compact objects and the early universe~\cite{2020-Aggarwal.etal-LivingRevRel}.
However, by leveraging the inverse Gertsenshtein effect (graviton-to-photon conversion), we can \emph{indirectly} constrain the properties of HFGWs by using radio telescopes~\cite{2021-Domcke.Garcia-Cely-PRL,Ito:2023nkq,2024-He.Sharma.etal-JCAP}, using  pulsars~\cite{Ito:2023fcr}, explaining the origin of fast radio bursts~\cite{2022-Kushwaha.etal-MNRAS,2023-Kushwaha.Sunil.Shanki-IJMPD} and constraining properties of evaporating PBHs~\cite{2020-Carr.etal-Rept.Prog.Phys}. Recently, in Ref.~\cite{2024-Jana.Shanki.etal-arXiv}, a gravitational analogue of the Gertsenshtein effect was proposed which might detect the solar mass black holes in the Milky Way galaxy.
	
GWs generated by parity-violating processes in the early universe are chiral, characterized by net non-zero circular polarization. In contrast, the standard model of cosmology predicts unpolarized GWs~\cite{1984-Kodama.Sasaki-PTPS,1992-Mukhanov.etal-Phy.Rep.,2009-Malik.Wands-PhysRept}. However, introducing modifications to the gravity sector, such as a time-dependent coupling to the Chern-Simons term, $f(\phi) R_{\mu\nu\alpha\beta}\tilde{R}^{\mu\nu\alpha\beta}$, induces gravitational birefringence, resulting in the production of helical GWs~\cite{1999-Lue.Wang.Kamionkowski-PRL,2003-Jackiw.Pi-PRD,2006-Alexander.Peskin.Jabbari-PRL,2009-Alexander.Yunes-PhyRept,2022-Peng.etal-PRD}. Additionally, parity-violating sources in the early universe, such as chiral fermions~\cite{2022-Gubler.etal-JCAP} and chiral gauge fields (with coupling $f(\phi) F_{\mu\nu} \tilde{F}^{\mu\nu}$)~\cite{2011-Sorbo-JCAP,2021-Okano.Fujita-JCAP}, also contribute to the generation of helical GWs. As mentioned earlier, due to their weak interaction with matter, GWs preserve information about the physical processes that produced them. The frequency of GWs encodes the characteristic length scale, while the amplitude reflects the energetics and efficiency of the source. Consequently, measuring the circular polarization of chiral GWs could provide insights into the parity-violating nature of these sources~\cite{2007-Saito.etal-JCAP,2010-Gluscevic.Kamionkowski-PRD,2013-Crowder.etal-PLB,2020-Domcke.etal-JCAP}.

In this work, we investigate the transfer of chirality from the \emph{EM sector} encompassing both the background magnetic field and propagating EM waves to the \emph{GW sector}. Specifically, we consider the photon-graviton conversion in the very early radiation-dominated (RD) epoch, wherein this mechanism is most efficient.
To the best of our knowledge, this study is the first to provide a detailed analysis of chirality transfer in photon-graviton conversion and to derive the relation estimating the net circular polarization between the two sectors. Moreover, we show that for a helical magnetic field, the conversion probability depends on the product of both the helicity components. This is one of the differentiating feature of this mechanism and is in contrast to other chiral GW production scenarios wherein the produced GWs are agnostic of the decaying helicity modes of the magnetic field~\cite{2011-Sorbo-JCAP,2021-Brandenburg.Sharma.etal-ApJ}.
In these works, the chirality parameter and polarization of the produced GWs, are solely determined by the dominant helicity mode of the helical magnetic field (i.e., either due to right-handed or left-handed mode), which is different in our case as the conversion probability (and hence, the chiral GWs spectrum) depends on the geometric mean $\sqrt{\bar{B}_+ \bar{B}_-}$, and not just on $\bar{B}_+$ or $\bar{B}_-$ (i.e., right-handed or left-handed helicity mode). 
For the primordial magnetic fields consistent with the Big-Bang Nucleosynthesis bound~\cite{2012-Kawasaki.Kusakabe-PRD,2016-Subramanian-Arxiv}, we calculate the energy density of the produced GWs contributing to the total stochastic GW background. Our results show that the (comoving) frequency of the GWs peaks $\sim 100 ~{\rm GHz}$ and is sensitive to the total relativistic degrees of freedom.  We also compare the characteristic GW strain ($h_c$) of the produced gravitons with other high-frequency GW sources and contrast our predictions with the sensitivity of current and upcoming missions aimed at detecting high-frequency GWs.

We follow the $(-,+,+,+)$ metric signature, Greek indices denote the spacetime and Latin indices refer to the spatial part only, and the reduced Planck mass is defined as $M_P^{-2} = 8\pi G$ where $M_P = 2.43\times 10^{18}\, {\rm GeV}$. We work with natural units where $\hbar = c = k_B = 1$. For the electromagnetic fields, we work in Lorentz-Heaviside units, where $1\, {\rm GeV}^2 = \sqrt{4\pi} \times 1.4\times 10^{19}\, {\rm Gauss}$.

\section{Photon-Graviton conversion in a background magnetic field}

In this section, we discuss the phenomenon of conversion of the EM waves to GWs and vice-versa~\cite{1962-Gertsenshtein-JETP,1974-Zeldovich-SJETP}.
To keep the analysis generic, we consider a system where the total action describing the photon-graviton oscillation involves the EM field coupled to the background spacetime and the scalar field, and is described by
\begin{equation}\label{total-action}
S_{\rm total} = \frac{1}{16\pi G} \int d^4 x \sqrt{-g} R - \int d^4 x \sqrt{-g} \left[  \frac{1}{2} \nabla_{\mu}\phi \nabla^{\mu}\phi + V(\phi)
\right] + \int d^4 x \sqrt{-g}  ~L_{\rm EM}
\end{equation}
where the first term is the Einstein-Hilbert action and the second term is the action for the (pseudo) scalar field. The third term describes the dynamics of the EM field whose Lagrangian is given by
\begin{align}\label{em-lagrangian}
L_{\rm EM} =   -\frac{1}{4} F_{\mu\nu} F^{\mu\nu} + J_{\mu} A^{\mu}  -   \frac{1}{4} f(\phi) F_{\mu\nu} \tilde{F}^{\mu\nu}  
\end{align}
where $J^{\mu}$ is the current source, $A^{\mu}$ is the EM vector field, $F_{\mu\nu} = \nabla_{\mu} A_{\nu}-\nabla_{\nu} A_{\mu}$ is the EM field tensor, and its dual is defined as $\tilde{F}^{\mu\nu} = \frac{\epsilon^{\mu\nu\alpha\beta} }{2} F_{\alpha\beta} = \epsilon^{\mu\nu\alpha\beta} \partial_{\alpha}A_{\beta}$. 
The last term in Eq.~\eqref{em-lagrangian} is a model-dependent term where the coupling function $f(\phi)$ determines the time evolution of the EM field~\cite{2011-Sorbo-JCAP}. 

To study the photon-graviton oscillation in the presence of a background magnetic field which is static and homogeneous, it is important to make the following distinctions
\begin{align}\label{EM_tensor_metric_tensor_pert}
 g_{\mu\nu} = \bar{g}_{\mu\nu} + h_{\mu\nu} , \quad g^{\mu\nu} = \bar{g}^{\mu\nu} - h^{\mu\nu} , \quad F_{\mu\nu} = \bar{F}_{\mu\nu} + \delta F_{\mu\nu}
\end{align}
where $\bar{g}_{\mu\nu}$ is the background metric and $h_{\mu\nu}$ are the fluctuations in the metric describing GWs with $ h_{\mu\nu} \ll |\bar{g}_{\mu\nu}|$ and $\text{det} (g_{\mu\nu}) = \text{det} (\bar{g}_{\mu\nu}) + \mathcal{O} (\text{det} (g_{\mu\nu}))$. The EM field tensor  $\bar{F}_{\mu\nu}$ refers to the background part which includes the contribution from the background magnetic field only, and $\delta F_{\mu\nu}$ refers to the propagating EM waves. It is important to note that while the covariant tensor $F_{\mu\nu}$ depends solely on the background magnetic field, its contravariant counterpart $F^{\mu\nu}$ couples with metric fluctuations as follows
\begin{align}\label{f_munu-pert}
F^{\mu\nu} = g^{\mu\alpha} g^{\nu\beta} F_{\alpha\beta} &= \bar{g}^{\mu\alpha} \bar{g}^{\nu\beta} \bar{F}_{\alpha\beta} + \left( \bar{g}^{\mu\alpha} \bar{g}^{\nu\beta} \delta F_{\alpha\beta} -  \bar{g}^{\mu\alpha} h^{\nu\beta} \bar{F}_{\alpha\beta}  - h^{\mu\alpha} \bar{g}^{\nu\beta}  \bar{F}_{\alpha\beta}  \right)
\nonumber \\
& + \left( h^{\mu\alpha} h^{\nu\beta} \bar{F}_{\alpha\beta} -  \bar{g}^{\mu\alpha} h^{\nu\beta} \delta F_{\alpha\beta}  - h^{\mu\alpha} \bar{g}^{\nu\beta}  \delta F_{\alpha\beta}  \right) + h^{\alpha\mu} h^{\beta\nu} \delta F_{\alpha\beta} , 
\end{align}
where background and the first-order terms are given by
\begin{align}\label{f_munu-first}
\bar{F}^{\mu\nu} = \bar{g}^{\mu\alpha} \bar{g}^{\nu\beta} \bar{F}_{\alpha\beta} , \quad \delta F^{ \mu\nu} =   \bar{g}^{\mu\alpha} \bar{g}^{\nu\beta} \delta F_{\alpha\beta} -  \left( \bar{g}^{\mu\alpha} h^{\nu\beta}  + h^{\mu\alpha} \bar{g}^{\nu\beta} \right)  \bar{F}_{\alpha\beta}  .
\end{align}
The evolution of the EM field is determined by the modified Maxwell equations, which can be derived by varying the action~\eqref{total-action} with respect to the gauge field $A_{\mu}$ as follows
\begin{align}\label{eom-covariant-full}
\partial_{\mu} \left[ \sqrt{-g}  \left\{ F^{\mu\nu} + f(\phi) \tilde{F}^{\mu\nu}  ~\right\} \right] + \sqrt{-g} J^{\nu} = 0  .
\end{align}
As we can see from the above equation, the choice of the coupling function $f(\phi)$ 
gives different propagation to the EM waves, for instance, one helicity mode enhances while the other decays. This phenomenon is called the \emph{birefringence effect}.
Now, the energy-momentum tensor for the EM field can be obtained by varying the Lagrangian \eqref{em-lagrangian} with respect to the metric $g^{\mu\nu}$, which is given by 
\begin{align}\label{em-tensor-standard}
T_{\mu\nu}^{\rm EM} = F_{\mu\alpha} {F_{\nu}}^{\alpha} - \frac{1}{4} g_{\mu\nu} F_{\alpha\beta} F^{\alpha\beta} - 2 A_{\mu} J_{\nu} + g_{\mu\nu} A_{\alpha} J^{\alpha}  .
\end{align}

Up to this point, we have derived the expressions for a general spacetime. However, to explore the cosmological or astrophysical implications of photon-graviton oscillations, it is essential to work within the framework of flat FRW spacetime with the line element
\begin{align}\label{frw-metric}
ds^2 = a^2 (\eta) [-d\eta^2 + \delta_{ij} dx^i dx^j]
\end{align}
where $a(\eta)$ is the scale factor and $\eta$ is conformal time which is related to cosmic time ($t$) by the relation $dt = a(\eta) d\eta$. 
The Maxwell's equations in FRW spacetime~\eqref{frw-metric} for the gauge field $A^i$ can be obtained from the action~\eqref{eom-covariant-full},
which in Coulomb gauge ($A^0 = 0, \partial_i A_i = 0$) are given by
\begin{align}\label{maxwell-eq-1}
\Box A_i + \eta_{ijl} f^{\prime} \partial_j A_l  = \frac{\bar{B}_k}{a^3} \eta_{k jl} \partial_j h_{li}
\end{align}
where $\Box \equiv -\frac{d^2}{d\eta^2} + \partial_i \partial_i$ is d'Alembertian operator in 4D. $\bar{\textbf{B}}$ and $\bar{\textbf{A}}$ are the background magnetic field and corresponding gauge field. The propagating EM waves' gauge field is denoted by $\textbf{A}$ and $\delta F_{ij} = \partial_i A_j - \partial_j A_i$.  In obtaining Eq.\eqref{maxwell-eq-1}, we used the following relations~\cite{2011-Durrer.Hollenstein.Jain-JCAP,2013-Durrer.Neronov-Arxiv,2016-Subramanian-Arxiv}
\begin{align}\label{relation-1}
    B_{\mu} = \epsilon_{\mu\nu\gamma} F^{\nu\gamma}/2 ~~, \qquad  a^{-4} \delta^{mj} \bar{F}_{jl} \partial_m h_{pi} = \bar{F}^{mp} \partial_m h_{pi} = \frac{\bar{B}_k}{a^3} \eta_{kmp} \partial_m h_{pi}
\end{align}
where $\eta^{0ijk} = \eta^{ijk} = \eta_{ijk}$ is the 3D totally antisymmetric symbol with $\eta_{123} = 1 = \eta^{123}$.

As we mentioned, for the photon-graviton oscillation, the background magnetic field  $\bar{\textbf{B}}$ should be transverse to the direction of propagation of EM waves (and GWs). 
Therefore, considering the propagating waves along the $z-$axis automatically fixes the background magnetic field to be in the $(x-y)$ plane. 
We will show later that, to study the chirality transfer for the following physical scenario: \emph{incoming wave} $+$ \emph{background magnetic field} $\rightarrow$ \emph{outgoing wave}, it is \emph{necessary} to consider the background magnetic field in the plane. Thus, we take the background magnetic field, $\bar{B}_{\mu} = a(\eta) (0,\bar{B}_1,\bar{B}_2,0)$. Furthermore, we can express the vector field in the Cartesian basis and the helicity basis as follows~\cite{2011-Durrer.Hollenstein.Jain-JCAP}
\begin{subequations}\label{vector-field-in-basis}
    \begin{equation}
        \textbf{A} = A_1 \epsilon_1 + A_2 \epsilon_2 = A_+ \epsilon_+ + A_- \epsilon_- 
    \end{equation}
    \begin{equation}
        \bar{\textbf{B}} = \bar{B}_1 \epsilon_1 + \bar{B}_2 \epsilon_2 = \bar{B}_+ \epsilon_+ + \bar{B}_- \epsilon_-
    \end{equation}
\end{subequations}
where $\epsilon_1 = \hat{x},\epsilon_2 = \hat{y}, \epsilon_3 = \hat{z}$ form the Cartesian basis and are related to the helicity basis as
\begin{align}\label{helicity-rel-1}
    \epsilon_{\pm} = \frac{\epsilon_1 \pm i \epsilon_2}{\sqrt{2}}  .
\end{align}
Using Eq.\eqref{vector-field-in-basis} and Eq.\eqref{helicity-rel-1}, one obtains the following relations for the gauge field and the magnetic field in the helicity basis
    \begin{equation}\label{helicity-relation-2}
        A_{\pm} = \frac{A_1 \mp i A_2}{\sqrt{2}} , \qquad B_{\pm} = \frac{B_1 \mp i B_2}{\sqrt{2}} ~~.
    \end{equation}
As we can see from the above relation, choosing the background magnetic field along one axis, for example, $\bar{\textbf{B}} = \bar{B}_1$ gives $\bar{B}_+ = \bar{B}_-$ would intrinsically assume that background magnetic field is non-helical. This suggests that the helicity information of the background magnetic field is lost. 
Therefore, to study the chirality transfer, it is necessary to consider the magnetic field to be in the $(x-y)$ plane. It should be noted that this requirement is not needed for non-chiral scenarios (see for example Refs.~\cite{2012-Dolgov.Ejlli-JCAP,2020-Fujita.Kamada.Nakai-PRD,2019-Domcke.etal-JCAP}). 
Given that $\bar{{\textbf{B}}}$ is in $(x-y)$ plane, using $\bar{\textbf{B}} = \nabla \times \bar{\textbf{A}}$, we can obtain the expression for the corresponding gauge field as $\bar{\textbf{A}} = \left( z \bar{B}_2, - z \bar{B}_1, 0  \right)$.
Now, following Ref.~\cite{Book-Carroll-GR}, the GW fluctuations in the helicity basis can be defined as
    \begin{equation}\label{helicity-gw-rel}
        h_{L} = \frac{h_+ - i h_{\times}}{\sqrt{2}}, \qquad h_{R} = \frac{h_+ + i h_{\times}}{\sqrt{2}} ~~.
    \end{equation}
Using Eq.\eqref{helicity-relation-2} and Eq.\eqref{helicity-gw-rel}, the Maxwell equation \eqref{maxwell-eq-1} in the helicity basis can be written as
\begin{subequations}\label{maxwell-eq-main-frw}
    \begin{equation}
        \Box A_+ - i f^{\prime} \partial_z A_+ = - \frac{i}{a^2} \bar{B}_- \partial_z h_L 
    \end{equation}
    \begin{equation}
        \Box A_- + i f^{\prime} \partial_z A_- =  \frac{i}{a^2} \bar{B}_+ \partial_z h_R
    \end{equation}
\end{subequations}
where the coupling of GW fluctuations with the background magnetic field $\bar{\textbf{B}}$ acts as a source for the propagating EM fields. From Eq.\eqref{maxwell-eq-main-frw}, GW and EM waves couple to each other with opposite helicity. This is because the mathematical structure of $A_{\sigma}, B_{\sigma}$ is similar to $h_{-\sigma}$, where $\sigma = +1$ (or $R$) is for right-handed and $\sigma = -1$ (or $L$) is for left-handed mode\footnote{In the literature the opposite definition of $h_{L,R}$ as compared to Eq.\eqref{helicity-gw-rel} is also used (see Ref.~\cite{2005-Alexander.Martin-PRD}), which will couple the GW and EM waves of same helicity. However, this will not affect the qualitative features of our results.}.

Before closing this section, we calculate the energy-momentum tensor for the EM fields, which for the source-free ($J^{\mu}=0$) region in FRW spacetime can be obtained from Eq.\eqref{em-tensor-standard} as
\begin{align}\label{em-tensor-expression}
    T_{ij}^{\rm EM} = - \left( E_i E_j + B_i B_j \right) + \frac{\delta_{ij} \delta^{lm} }{2} \left( E_l E_m + B_l B_m \right)
\end{align}
where we have used $F_{0i} = E_i = -\frac{1}{a (\eta)} A_i^{\prime} $ and $F_{ij} = -a^3 \eta_{ijk} B^k $.

\section{Chirality transfer in the Early Universe}

In the previous section, we obtained the equations of motion describing the photon-graviton oscillation in FRW spacetime. In this section, we particularly focus on the transfer of chirality from EM waves or background magnetic fields to GWs in the early RD epoch when the temperature of the universe was very high.
For waves with frequency $\omega$ larger than the Hubble parameter $H = \dot{a}/a$ (which characterizes the time scale of background expansion), we can ignore the cosmic expansion of the universe.
In that case, we can work with the flat Minkowski spacetime and the Maxwell equations can be obtained from Eq.\eqref{maxwell-eq-main-frw} as follows
\begin{subequations}\label{maxwell-eq-main}
    \begin{equation}
        \Box A_+ - i f^{\prime} \partial_z A_+ = - i\bar{B}_- \partial_z h_L 
    \end{equation}
    \begin{equation}
        \Box A_- + i f^{\prime} \partial_z A_- =  i\bar{B}_+ \partial_z h_R ~~.
    \end{equation} 
\end{subequations}
After obtaining these equations, we now focus on understanding the evolution of GWs. Following Ref.~\cite{2018-Caprini.Figueroa-CQG},  the equation of motion for the metric perturbations $h_{ij}$ in TT gauge ($\partial_i h^i_j = 0 = h^i_i,  h_{00} = 0 = h_{0i}$), in Minkowski spacetime is given by
\begin{align}\label{h_ij-TT-gauge}
\left( \partial_t^2 - \nabla^2  \right) h_{ij} = 16 \pi G \delta T^{TT}_{ij} ~~, 
\end{align}
where $T_{ij}^{TT}$ is the energy-momentum tensor of the EM fields in the TT gauge. We can obtain the transverse-traceless part of the energy-momentum tensor in Minkowski spacetime with the relation
\begin{align}\label{em-tensor-TT-gauge}
    T_{ij}^{TT} (\textbf{k}) = \sum_{m,l} \Lambda_{ij,ml} (\textbf{k}) T_{ml}^{\rm EM} (\textbf{k})
\end{align}
where $\Lambda_{ij,ml} (\textbf{k})$ is the projection operator~\cite{2018-Caprini.Figueroa-CQG,2020-Fujita.Kamada.Nakai-PRD} given by
\begin{align}\label{projection-tensor}
    \Lambda_{ij,ml} (\textbf{k}) = P_{im} (\textbf{k}) P_{jl} (\textbf{k}) - \frac{1}{2} P_{ij} (\textbf{k}) P_{ml} (\textbf{k}), \quad \text{where} \quad P_{ij} (\textbf{k}) = \delta_{ij} - \frac{k_i k_j}{|\textbf{k}|^2}  ~~.
\end{align}
Thus, using Eq.~\eqref{em-tensor-expression} in the flat Minkowski spacetime, we can compute perturbations in the energy-momentum tensor $T_{ij}$ in TT gauge as
\begin{subequations}\label{EMT-ttgauge}
    \begin{equation}
        \delta T_{xx}^{TT} = - \delta T_{yy}^{TT} = \bar{B}_y \delta B_y - \bar{B}_x \delta B_x 
    \end{equation}
    \begin{equation}
 \delta T_{xy}^{TT} =  \delta T_{yx}^{TT} = - ( \bar{B}_x \delta B_y + \bar{B}_y \delta B_x ) ~~.
    \end{equation}
\end{subequations}
Using Eq.\eqref{helicity-relation-2} in Eq.\eqref{h_ij-TT-gauge}, we can obtain the evolution of metric fluctuations sourced by the EM fields in helicity basis as follows
\begin{subequations}\label{gw-eq-main}
    \begin{equation}
        \left( \partial_t^2 - \nabla^2  \right) h_R = -16 \pi G i \bar{B}_- \partial_z A_-
    \end{equation}
    \begin{equation}
        \left( \partial_t^2 - \nabla^2  \right) h_L = 16 \pi G i  \bar{B}_+ \partial_z A_+
    \end{equation}
\end{subequations}
where $\delta B_- = i\partial_z A_-$ and $\delta B_+ = -i\partial_z A_+$. From these equations, we notice that in the photon-graviton conversion, the chirality in gravitons can be induced (or transferred) due to either the chiral nature of the background magnetic field or the chirality of the propagating EM waves. Therefore, this allows us to divide the analysis into two cases based on the chiral nature of the background magnetic field: (i) \emph{non-helical} background magnetic field (with helical propagating EM waves), and (ii) \emph{helical} background magnetic field (with either helical or non-helical propagating EM waves). In what follows, we will investigate these cases in more detail.

\subsection{\emph{Non-helical} background magnetic field $\bar{\textbf{B}}$}
\label{subsec:non-helical-B}

Let us first explore the simpler case where the background magnetic field is \emph{non-helical}, i.e., $\bar{B}_+ = \bar{B}_- = \bar{B}$. However, to study the chirality transfer to the GWs, the propagating EM waves must be helical and we consider them to be maximally helical with dominating right-handed mode, i.e., $|A_+| \gg |A_-|$\footnote{Note that, choosing the dominant left-handed mode $A_-$ will not change the amplitude of the conversion probability because the eigenvalues of matrix $\mathcal{M}$ \eqref{mixing-matrix-def} are quadratic in $\bar{\textbf{B}}$. The only effect would be to produce GWs of opposite helicity, i.e. right-handed GWs.}. As mentioned earlier, we focus only on the model-independent analysis, thus setting $f(\phi) = 0$. From Eq.\eqref{maxwell-eq-main} and \eqref{gw-eq-main}, we have
\begin{subequations}\label{case1-em-gw-1}
    \begin{equation}\label{case1-em-1}
        (\partial_t^2 - \nabla^2 + m_{\gamma}^2 ) A_+  = i \bar{B} \partial_z h_L
    \end{equation}
    \begin{equation}\label{case1-gw-1}
         (\partial_t^2 - \nabla^2)  h_L =   i 16 \pi G  \bar{B}  \partial_z A_+ 
    \end{equation}
\end{subequations}
where we introduced the effective mass of the photon $m_{\gamma}^2$ in Eq.\eqref{case1-em-1}. This effective mass arises due to the interaction of photons with the highly conducting plasma in the early Universe, which changes the dispersion relation between photon momentum ($k$) and its frequency ($\omega$) giving an effective mass to the photon, $m_{\gamma}^2 = \omega^2 - k^2 \neq 0$. 
To solve Eq.\eqref{case1-em-gw-1}, we follow the approach of Refs.~\cite{2020-Fujita.Kamada.Nakai-PRD,2012-Dolgov.Ejlli-JCAP,2013-Dolgov.Ejlli-PRD,2023-Dolgov.etal-Universe}. Focussing on the mode with specific (single) frequency of the propagating waves allows us to decompose the vectors and tensors as
\begin{align}\label{fourier-decomp}
    h_{\sigma} (z,t) = h_{\sigma} (z) e^{-i\omega t}, \qquad A_{\sigma} (z,t) = A_{\sigma} (z) e^{-i\omega t}
\end{align}
where we have separated the oscillating part (in time) of the wave from the spatial. Using Eq.\eqref{fourier-decomp} in \eqref{case1-em-gw-1} gives
\begin{subequations}\label{case1-em-gw-system-main}
		\begin{alignat}{2}
        \left( \omega^2 + \partial_z^2 - m_{\gamma}^2 \right) A_+ &= - i \bar{B} \partial_z h_L
        \\
        \left( \omega^2 + \partial_z^2 \right) h_L &= - \frac{2 i}{M_P^2} \bar{B} \partial_z A_+ ~~.
       	\end{alignat}
\end{subequations}
The above equations might look simpler but are not straightforward to solve exactly.
 However, they can be simplified by assuming that the background magnetic field varies in space on scales much larger than the photon wavelength. This allows us to perform the following expansion~\cite{2001-Deffayet.etal.Zaldarriaga-PRD}
 \begin{align}\label{omega-z-rel}
     \omega^2 + \partial_z^2 = (\omega + i\partial_z) (\omega - i \partial_z) \simeq 2\omega (\omega + i\partial_z )
 \end{align}
 where $-i\partial_z = k$ is the momentum operator, and assuming a general dispersion relation of the form $k=n\omega$ where $n$ is refractive index satisfying $|1-n| \ll 1$, we can write $\omega-i\partial_z = \omega + k \simeq 2\omega$. With this, we can write Eq.\eqref{case1-em-gw-system-main} in a simplified form as 
\begin{subequations}\label{case1-em-gw-simplified}
	\begin{alignat}{2}
				  \left( \omega + i\partial_z - \frac{m_{\gamma}^2}{2\omega} \right) A_+ & \simeq  \frac{\bar{B} }{\sqrt{2} M_P} \tilde{h}_L
		\\
		\left( \omega + i \partial_z \right) \tilde{h}_L & \simeq  \frac{\bar{B}}{\sqrt{2} M_P}  A_+ ~~.
	\end{alignat}
\end{subequations}
To make the above equations symmetric in $1/M_P$, we used a dimensionful quantity $\tilde{h}_{L/R} = \frac{M_p}{\sqrt{2}}h_{L/R}$ which has the same dimensions as that of the gauge field. These coupled equations for the photon-graviton oscillations can be solved in the same way as for the neutrino oscillations~\cite{2012-Dolgov.Ejlli-JCAP,2020-Fujita.Kamada.Nakai-PRD}. Thus, we can write the Eq.\eqref{case1-em-gw-simplified} in the matrix form
\begin{align}\label{oscillation-eq-matrix}
	\left[ (\omega + i\partial_z) \mathbb{I} + \mathcal{M} \right]   \hat{\Psi} (z)  \simeq 0
\end{align}
where $\mathbb{I}$ is the identity matrix, $\mathcal{M}$ is analogous to mixing matrix and $\hat{\Psi} (z)$ is the column vector field, given by
\begin{align}\label{mixing-matrix-def}
  \mathbb{I} = \begin{bmatrix}
    1 & 0 \\ 0 & 1
\end{bmatrix}, \quad  \mathcal{M} = \begin{bmatrix}
        -\frac{m_{\gamma}^2 }{2\omega} & -\frac{\bar{B}}{\sqrt{2} M_P} \\
        -\frac{\bar{B}}{\sqrt{2} M_P} & 0
    \end{bmatrix},    \quad \text{and} \quad
      \hat{\Psi} (z) = \begin{bmatrix}
    A_+ (z) \\ \tilde{h}_L (z)
    \end{bmatrix} ~ ~ .
\end{align}
To solve Eq.\eqref{oscillation-eq-matrix}, we use the Unitary matrix given by
\begin{align}\label{unitary-operator}
	\textbf{U} =   \begin{pmatrix}
		\cos \theta  & \sin \theta \\
		-\sin\theta & \cos\theta
	\end{pmatrix} \quad \text{and} \quad   \textbf{U}^T =   \begin{pmatrix}
		\cos \theta  & -\sin \theta \\
		\sin\theta & \cos\theta
	\end{pmatrix} ~~ ,
\end{align}
satisfying $\textbf{U} \textbf{U}^T = \textbf{U}^T \textbf{U} = \mathbb{I}$, which diagonalizes the mixing matrix $\mathcal{M}$ as follows
\begin{align}
    \textbf{U} \mathcal{M} \textbf{U}^T = \begin{bmatrix}
        \lambda_1 & 0 \\ 0 & \lambda_2
    \end{bmatrix} 
\end{align}
where $\lambda_{1,2}$ are the eigenvalues of $\mathcal{M}$, which are given by
\begin{align}\label{lambda-1-2}
    \lambda_{1,2} = - \frac{1}{2} \left[ \frac{m_{\gamma}^2 }{2\omega} \pm \sqrt{\left( \frac{m_{\gamma}^2}{2\omega} \right)^2 + \frac{2 {\bar{B}}^2}{M_P^2}} \right] ~~.
\end{align}
Let us now discuss the solutions of the Eq.\eqref{oscillation-eq-matrix}, which are given by the Unitary transformation of the field $\hat{\Psi} (z)$ as 
\begin{align}
	\hat{\Psi}'(z) = \textbf{U} \hat{\Psi} (z) = \exp \left\{i \int_0^z (\omega + \tilde{\mathcal{M}}) dz \right\} \hat{\Psi}'(0)
\end{align}
where $\tilde{\mathcal{M} } = \mathbf{U} \mathcal{M} \mathbf{U}^T$.
Now multiplying from the left-hand side by $\textbf{U}^T$ in the above equation gives
\begin{align}\label{psi-equation}
  \Psi (z) =  \textbf{U}^T \hat{\Psi}'(z) = \exp \left\{i \int_0^z (\textbf{U}^T \tilde{\mathcal{M}} \textbf{U} ) dz \right\} \cdot \exp \left\{i \int_0^z \omega dz \right\} \hat{\Psi}(0)
\end{align}
where $e^{i \int_0^z \omega dz }$ is a phase factor and can be absorbed in $\hat{\Psi} (0)$. Furthermore, using Eq.\eqref{unitary-operator}, we can calculate the first term as
\begin{align}
   e^{i \int_0^z \textbf{U}^T \tilde{\mathcal{M}} \textbf{U} dz } 
   = \begin{pmatrix}
        \cos^2 \theta ~ e^{i \lambda_1 |z|} + \sin^2 \theta ~ e^{i \lambda_2 |z|} & \quad \left( e^{ i\lambda_1 |z|} - e^{i \lambda_2 |z|} \right) \sin\theta \cos\theta\\
        \left( e^{i \lambda_1 |z|} - e^{i \lambda_2 |z|} \right) \sin\theta \cos\theta & \quad\sin^2 \theta ~ e^{i \lambda_1 |z|} + \cos^2 \theta ~ e^{i \lambda_2 |z|} 
    \end{pmatrix} ~~.
\end{align}
Thus, using the above result in Eq.\eqref{psi-equation} and comparing the components of the column vector on both sides gives the solution of the photon-graviton coupled equation \eqref{oscillation-eq-matrix} as
\begin{subequations}\label{photon-graviton-solution-main}
	\begin{alignat}{2}
		    A_+ (z) &\simeq \left( \cos^2 \theta ~ e^{i \lambda_1 |z|} + \sin^2 \theta ~ e^{i \lambda_2 |z|} \right) A_+ (0) + \left( \left( e^{ i\lambda_1 |z|} - e^{i \lambda_2 |z|} \right) \sin\theta \cos\theta \right) \tilde{h}_L (0)
		\\
		\tilde{h}_L (z) &\simeq  \left( \left( e^{ i\lambda_1 |z|} - e^{i \lambda_2 |z|} \right) \sin\theta \cos\theta \right) A_+ (0) +  \left( \sin^2 \theta ~ e^{i \lambda_1 |z|} + \cos^2 \theta ~ e^{i \lambda_2 |z|} \right) \tilde{h}_L (0) ~~.
	\end{alignat}
\end{subequations}
From the above equation, we can easily calculate the conversion probability of photons to gravitons. Assuming there are only photons at the beginning and \emph{no} gravitons provides the following initial conditions: $A_+ (0) = 1$ and $\tilde{h}_L (0) = 0$, which gives the following relation
\begin{align}\label{conv-probability-1}
    \left| \langle A_+ (0) | \tilde{h}_L (z) \rangle \right|^2 = \sin^2\theta \cos^2\theta  \left| e^{ i\lambda_1 |z|} - e^{i \lambda_2 |z|} \right|^2
    = \sin^2 (2\theta) ~ \sin^2\left( \frac{(\lambda_1 - \lambda_2) |z|}{2} \right)
\end{align}
where the mixing angle $\theta$ is the ratio of the off-diagonal term to the difference of the diagonal terms~\cite{1988-Raffelt.Stodolsky-PRD}, and is given by $\tan(2\theta) = \frac{2\sqrt{2} \omega \bar{B} }{m_{\gamma}^2 M_P} $. Therefore, the conversion probability can be estimated by
using Eq.\eqref{lambda-1-2} in Eq.\eqref{conv-probability-1} as
\begin{align}\label{conv-probability-2}
 \mathcal{P}_{\gamma \rightarrow g} (\ell) =   \left| \langle A_+ (0) | \tilde{h}_L (z) \rangle \right|^2
    =\left( \frac{8\omega^2 {\bar{B}}^2}{m_{\gamma}^4 M_P^2 + 8 \omega^2 {\bar{B}}^2} \right) ~ \sin^2\left[ \frac{|z|}{2} \sqrt{ \left( \frac{m_{\gamma}^2}{2\omega} \right)^2 + \frac{2 {\bar{B}}^2}{M_P^2} } \right] ~~.
\end{align}
where $|z| = \ell$ is the typical length of the photon propagation and also the length scale over which the conversion happens. Note that if this takes place in the early RD epoch when the temperature of the universe is very high, $T \gg m_e$, the effective photon mass is dominated by the Debye mass with $m_{\gamma}^2 = m^2_D \simeq e^2 T^2$ where $e = \sqrt{4\pi \alpha_e} \approx 0.3$, $m_e \simeq 0.51 \, {\rm MeV}$ is the mass of electron, and $\alpha_e$ is the fine structure constant. Furthermore, during this epoch, we can make some realistic assumptions to simplify Eq.\eqref{conv-probability-2}. 
For example, considering the background magnetic fields to be relatively larger with $\bar{B} \lesssim T^2$ and as long as $\omega \ll M_P$, we have $\frac{ \bar{B}}{M_P} \ll  \frac{m_{\gamma}^2}{2\omega}  $. Under these approximations, we obtain
\begin{align}
     \frac{8\omega^2 {\bar{B}}^2}{m_{\gamma}^4 M_P^2 + 8 \omega^2 {\bar{B}}^2}  \simeq \frac{8\omega^2 {\bar{B}}^2}{m_{\gamma}^4 M_P^2 }, \quad \text{and} \quad
     \frac{\ell}{2} \sqrt{ \left( \frac{m_{\gamma}^2}{2\omega} \right)^2 + \frac{2 {\bar{B}}^2}{M_P^2} } \simeq \frac{1}{2\Gamma_{\gamma}}\frac{m_{\gamma}^2}{2\omega} \simeq \frac{\pi T}{\alpha_e \omega} ~~.
\end{align}
where we used the fact that $\ell$ for a wave packet of the photon is equal to the mean free path of the photons, $\ell \simeq \Gamma_{\gamma}^{-1}$, where $\Gamma_{\gamma} \simeq \alpha_e^2 T$ is the scattering rate of the photon with charged particles in the thermal bath. 
Within these assumptions, and for the case where $ \alpha_e \omega  \ll \pi T $, we can calculate the ensemble average of Eq.\eqref{conv-probability-2} by approximating the sinusoidal term by $1/2$, which gives the probability of conversion\footnote{Note that, the other assumption where $ \alpha_e \omega  \gg \pi T $ would give $\sin\left( \frac{\pi T}{\alpha_e \omega}\right) \sim \frac{\pi T}{\alpha_e \omega} \ll 1$ and the conversion probability would be highly suppressed by a very small factor $\sim \left( \frac{\pi T}{\alpha_e \omega}\right)^2$ as compared to the one obtained in Eq.\eqref{conv-probability-final}, hence we will not consider this regime.}
\begin{align}\label{conv-probability-final}
   \langle  \mathcal{P}_{\gamma \rightarrow g} (\ell) \rangle
    \simeq \frac{4 \omega^2 {\bar{B}}^2}{m_{\gamma}^4 M_P^2 }  \simeq \frac{4 \omega^2 {\bar{B}}^2}{e^4 T^4 M_P^2 }  ~~.
\end{align}
\begin{figure}[t]
\centering
\includegraphics[height=2.8in]{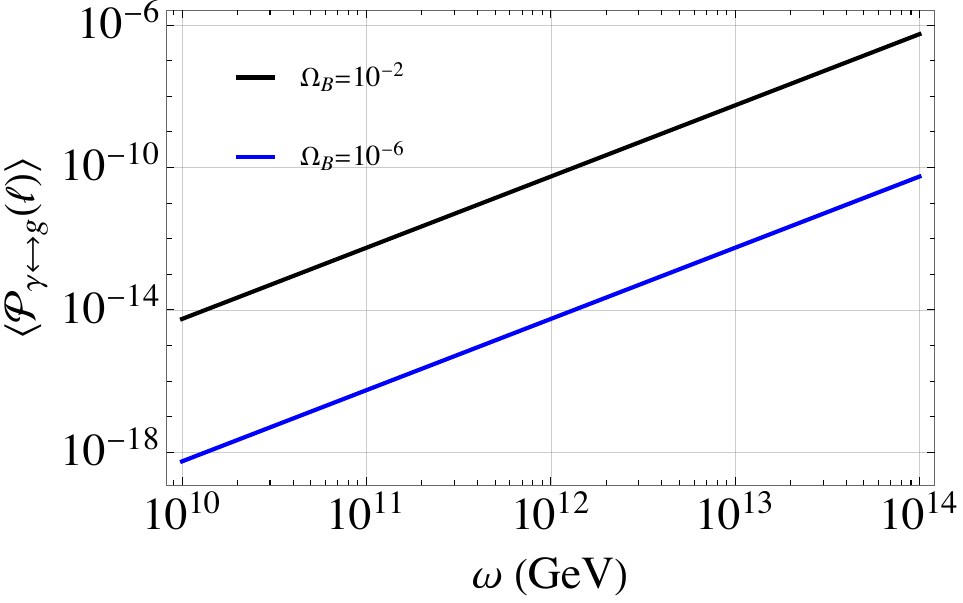} 
\caption{Ensemble averaged conversion probability for the gravitons (left-handed mode), which increases quadratically with $\omega$ for a fixed $\Omega_B$.}
\label{fig:conv-probability}
\end{figure}

On superhorizon scales, the magnetic fields evolve adiabatically, $\bar{B} \propto a^{-2}$; however, this evolution behaviour is highly affected by magnetohydrodynamics (MHD) effects on subhorizon scales. Therefore, rather than working with the strength of (comoving) background magnetic field, it is more convenient to use the fractional density parameter $\Omega_B$ defined as the ratio of magnetic energy density to the total energy density of the Universe, defined by
\begin{align}
	\Omega_B (T) \equiv \frac{\rho_B (T)}{\rho_{\rm tot} (T)} = \frac{15 {\bar{B}}^2 (T) }{\pi^2 g_* (T) T^4} 
\end{align}
where $\rho_{\rm tot} (T) = \pi^2 g_* (T) T^4/30$ is the total energy density of the Universe at temperature $T$ during the RD epoch. 
In order to preserve the isotropy of the background FRW Universe, one must impose the condition that the energy density in the magnetic fields at a given epoch remains subdominant, which is characterized by the condition $\Omega_B (T) <1$.
In terms of $\Omega_B$, the conversion probability in \eqref{conv-probability-final} is given by
\begin{align}\label{conversion-probability-OmegaB}
	\langle \mathcal{P}_{\gamma \leftrightarrow g} (\ell)  \rangle & \simeq \frac{4 \pi^2 g_* (T) \omega^2 }{15 e^4 M_P^2 } \Omega_B (T) \nonumber 
	\\
	& \simeq 5.28\times10^{-5} \left( \frac{g_* (T)}{100} \right) \left( \frac{\omega}{10^{14} GeV} \right)^2 \Omega_B (T) ~~ .
\end{align}
\ref{fig:conv-probability} shows the behavior of Eq.\eqref{conversion-probability-OmegaB} for different values of $\Omega_B$ and suggests that the conversion probability is higher for the stronger background magnetic field. This is expected because the stronger the catalyst (background magnetic field), the stronger would be the effect.

\subsection{\emph{Helical} background magnetic field $\bar{\textbf{B}}$}
\label{subsec:helical-background-B}

Now, let us discuss the more generic case where the background magnetic field is \emph{helical}. However, the propagating EM waves could be helical or non-helical. Following the analysis in the previous case, from Eq.\eqref{maxwell-eq-main} and Eq.\eqref{gw-eq-main}, the equations for photon-graviton oscillations for the modes $(A_+ \leftrightarrow \tilde{h}_L)$ are given by
\begin{subequations}\label{case2-em-gw-simplified-Aplus}
    \begin{alignat}{2}
	\left( \omega + i\partial_z - \frac{m_{\gamma}^2}{2\omega} \right) A_+ &\simeq  \frac{\bar{B}_- }{\sqrt{2} M_P} \tilde{h}_L	\\
 \left( \omega + i \partial_z \right) \tilde{h}_L & \simeq  \frac{\bar{B}_+}{\sqrt{2} M_P}  A_+
    \end{alignat}
\end{subequations}
and for the modes $(A_- \leftrightarrow \tilde{h}_R)$ are given by
\begin{subequations}\label{case2-em-gw-simplified-Aminus}
    \begin{alignat}{2}
 \left( \omega + i\partial_z - \frac{m_{\gamma}^2}{2\omega} \right) A_- &\simeq  -\frac{\bar{B}_+ }{\sqrt{2} M_P} \tilde{h}_R	\\
	\left( \omega + i \partial_z \right) \tilde{h}_R & \simeq  - \frac{\bar{B}_-}{\sqrt{2} M_P}  A_- ~~.
    \end{alignat}
\end{subequations}
We employ the same approximations and assumptions as in the previous case, which allows us to make a consistent comparison between the two cases. However, as compared to Eq.\eqref{case1-em-gw-simplified}, we should notice an important difference in the RHS of Eq.~\eqref{case2-em-gw-simplified-Aplus} and Eq.\eqref{case2-em-gw-simplified-Aminus} where the helicity of the background magnetic field enters as a source driving the conversion mechanism. Furthermore, we can construct the mixing matrix $\mathcal{M}$ for Eq.~\eqref{case2-em-gw-simplified-Aplus} and Eq.\eqref{case2-em-gw-simplified-Aminus}, where the off-diagonal terms contain both helicity components of the background magnetic field. Note that the mixing angle and the eigenvalues for both are same and are given by $\tan (2\theta) = \frac{2\sqrt{2} \omega \sqrt{\bar{B}_+ \bar{B}_-}}{m_{\gamma}^2 M_P}$ and $ \lambda_{1,2} = - \frac{1}{2} \left[ \frac{m_{\gamma}^2 }{2\omega} \pm \sqrt{\left( \frac{m_{\gamma}^2}{2\omega} \right)^2 + \frac{2 \bar{B}_+ \bar{B}_- }{M_P^2}} \right] $, respectively. Assuming \emph{no} gravitons of either helicity initially, i.e., $\tilde{h}_L (0) = 0 = \tilde{h}_R$ (0) and only photons, allows us to obtain the following 
\begin{subequations}
    \label{case2-conv-probability-1}
\begin{alignat}{2}
    \left| \langle A_+ (0) | \tilde{h}_L (z) \rangle \right|^2 
    &= \sin^2 (2\theta) ~ \sin^2\left( \frac{(\lambda_1 - \lambda_2) |z|}{2} \right) ~ |\langle A_+ (0) | A_+ (0) \rangle |^2
    \\
    \left| \langle A_- (0) | \tilde{h}_R (z) \rangle \right|^2 
    &= \sin^2 (2\theta) ~ \sin^2\left( \frac{(\lambda_1 - \lambda_2) |z|}{2} \right) ~ \delta_A^2 ~ |\langle A_+ (0) |  A_+ (0) \rangle |^2
\end{alignat}
\end{subequations}
where we have defined $\sqrt{\delta_A}$ as $ |A_-| = \sqrt{\delta_A} \, |A_+|$. As we can see from the RHS of the above equation, the term $|\langle A_+ (0) |  A_+ (0) \rangle |$ is an overall normalization that can be set to unity without any loss of generality. 
Therefore, following Eq.~\eqref{conversion-probability-OmegaB}, the conversion probability for both the helicity modes can be estimated as
\begin{subequations}\label{case2-conversion-probability}
    \begin{alignat}{2}
         \langle \mathcal{P}^L_{\gamma \rightarrow g} (\ell) \rangle 
    &= \frac{4\pi^2 g_* (T)}{15 e^4} \left( \frac{\omega}{M_P}\right)^2 \, \, \sqrt{\delta_B} \, \Omega_B^+
    \\
    \langle \mathcal{P}^R_{\gamma \rightarrow g} (\ell) \rangle 
    &=\frac{4\pi^2 g_* (T)}{15 e^4} \left( \frac{\omega}{M_P}\right)^2 \delta_A^2 \, \, \sqrt{\delta_B} \, \Omega_B^+
    \end{alignat}
\end{subequations}
where we defined $|\bar{B}_-| = \sqrt{\delta_B} \, |\bar{B}_+ | $ and $\Omega_B^{\pm} = \frac{15 \bar{B}_{\pm}^2}{\pi^2 g_* T^4}$. 
Considering the maximally helical gauge field with $\delta_A \ll 1$ (i,e., $|A_-| \ll |A_+|$) implies $\langle \mathcal{P}^L_{\gamma \rightarrow g} \rangle \gg \langle \mathcal{P}^R_{\gamma \rightarrow g} \rangle $ which is consistent with our previous result in Eq.\eqref{conv-probability-final} in the limit $\delta_B \rightarrow 1$ (i.e., non-helical background magnetic field).

Before we close this section, we would like to make a few important remarks about the results obtained in Eq.\eqref{case2-conversion-probability}. It shows that irrespective of its helical nature, the background magnetic field affects both left- and right-handed GWs in a similar way due to the term $\sqrt{\delta_B} \Omega_B^+ \sim \bar{B}_+ \bar{B}_-$. However, the product of both helicity modes of the background magnetic field depends on their relative amplitude with the condition $\sqrt{\delta_B} \Omega_B^+ < 1$, to avoid a magnetic field dominant Universe. In most of the inflationary magnetogenesis scenarios (even with a non-trivial reheating epoch), both modes evolve in a similar fashion about their geometric mean (i.e.,$\sqrt{\bar{B}_+ \bar{B}_-}$), see Refs.\cite{2011-Durrer.Hollenstein.Jain-JCAP,2019-Fujita.Durrer-JCAP}. 
If the effective field strength determined by $ \sim \sqrt{\bar{B}_+ \bar{B}_-}$ is strong enough, these scenarios can lead to significant conversion probability. Therefore, in contrast to other generation mechanisms wherein due to the addition of the contribution from both helicity modes (such as gauge field induced GWs~\cite{2011-Sorbo-JCAP}), the generated GWs are agnostic about the subdominant (decaying) helical mode of the magnetic fields. 

\section{The polarized gravitational wave background}
\label{sec:gwb-calculation}

In this section, we estimate the energy spectrum of the produced gravitons, which would determine their contribution to the universe's total stochastic GW background.
The energy spectrum can be calculated using the Boltzmann equation which describes the evolution of the distribution function of GWs sourced by thermal photons at high temperature in the early Universe~\cite{2020-Fujita.Kamada.Nakai-PRD,2021-Domcke.Garcia-Cely-PRL}
\begin{equation}\label{boltzmannEq-T}
	- H \left( T \partial_T + \omega \partial_{\omega}  \right) f_g (T,\omega) = \Gamma_{\gamma \rightarrow g} f_{\gamma} (T,\omega )
\end{equation}
where $\Gamma_{\gamma \rightarrow g}$ is the conversion rate of thermal photons to gravitons per unit time which can be evaluated in terms of ensemble average of the conversion probability as
\begin{align}\label{Gamma-to-conv-probability}
	\Gamma_{\gamma \rightarrow g} = \frac{1}{2} \Gamma_{\gamma} \langle \mathcal{P}_{\gamma \rightarrow g} (\ell) \rangle \simeq \frac{ {\bar{B}}^2 \omega^2 }{8\pi^2 M_P^2 T^3} ~~.
\end{align}
Since the distribution function of thermal photons is determined by the Bose-Einstein distribution, $f_{\gamma} (T,\omega) = \frac{1}{e^{\omega/T} - 1 }$, using Eq.\eqref{Gamma-to-conv-probability} in the Boltzmann Equation \eqref{boltzmannEq-T} gives
\begin{align}\label{boltzman-eq-main}
	\left( T \partial_T + \omega \partial_{\omega}  \right) f_g (T,\omega) \simeq - \left(\frac{g_* (T) T_{\rm end}^2}{120 M_P^2 H_{\rm end}} \Omega_B (T) \right)\frac{\omega^2}{T} \frac{1}{e^{\omega/T} - 1} ~~.
\end{align}
In obtaining the above relation, we used $H = H_{\rm end} \frac{T^2}{T_{\rm end}^2}$ during RD epoch where $`$end' refers to the quantities evaluated at the end of inflation (which we take as the onset of RD epoch). Note that during the RD era where the conversion mechanism takes place, the term in the bracket on the RHS has a weak dependence on $T$ (as compared to other terms outside the bracket) and can be approximated as a constant, which allows us to solve the above equation approximately analytically. Using the boundary condition that there are no gravitons produced from this mechanism before the onset of the RD epoch, $f_g (T_{\rm end},\omega) = 0$, the solution of the above equation can be given by
\begin{align}\label{fg-final}
	f_g (T,\omega) \simeq \frac{g_* (T) T_{\rm end}^2}{120 M_P^2 H_{\rm end}} \Omega_B  \frac{\omega^2}{T^2} \frac{ ( T_{\rm end} - T)}{e^{\omega/T} - 1} ~~.
\end{align}
The above distribution function of the produced gravitons also allows us to compare the peak frequencies of the intensities of these gravitons with that of thermal photons by using the intensity ($I$) relation, $I \propto \omega^3 f(\omega)$, which gives
\begin{subequations}\label{intensity-frequency}
	\begin{alignat}{2}
		I_{\gamma} &\propto \frac{\omega^3}{e^{\omega/T}-1} \quad \implies \quad \omega_{\gamma}^{\rm peak} = 2.82 T_{\gamma}
		\\
		I_g & \propto \frac{\omega^5}{e^{\omega/T}-1} \quad \implies \quad \omega_g^{\rm peak} = 4.965 T_g ~~ .
	\end{alignat}
\end{subequations}
where subscripts $g$ and ${\gamma}$ refer to graviton and photon, respectively. From the entropy conservation in the matter sector, we have $T_{\gamma} \propto g_{s}^{-1/3} a^{-1}$ while weakly interacting nature of gravitons with matter gives $T_{g} \propto a^{-1}$. Assuming during the photon-graviton conversion process both are in thermal equilibrium, and thus $T_{\gamma} = T_g$ at the beginning. However, due to the different decaying behaviour, the temperature of photons and gravitons at the present epoch are related as
\begin{align}\label{Tg-Tgamma-relation}
    T_g (t_0) = T_{\gamma} (t_0) \left( \frac{g_{s,0}}{g_s (t)} \right)^{1/3} ~~.
\end{align}
where $g_{s,0} = 3.93$ and $ T_{\gamma}(t_0) = 2.726 {\rm K} = 3.54\times 10^2 \, {\rm GHz} $ is the temperature of the photons at present epoch~\cite{Book-Dodelson-2020}. Using Eq.~\eqref{intensity-frequency} with Eq.~\eqref{Tg-Tgamma-relation} gives the relation between the peak frequencies at the present epoch as $\omega_g^{\rm peak} (t_0) = 2.77 \, 
g_s^{-1/3} (t) \omega_{\gamma}^{\rm peak} (t_0)$. Now let us compute the energy spectrum of the produced gravitons which would contribute to the total stochastic GW background along with other sources in the early Universe. This background is characterized by the quantity $\Omega_{GW}$ which parametrizes the energy density of gravitons per logarithmic frequency bin at present epoch~\cite{2020-Fujita.Kamada.Nakai-PRD,2021-Domcke.Garcia-Cely-PRL}
\begin{align}\label{gwb-definition}
	\Omega_{GW,0} \equiv \frac{1}{\rho^0_c} \frac{\rho_{GW} (\omega,t_0)}{d \ln \omega}, \qquad \text{where} \quad \rho_{GW} (\omega) \equiv \int \frac{d\ln\omega ~ }{\pi^2} \omega^4 f_g (\omega)
\end{align}
where the critical density of the universe at present epoch $\rho_c^0 = \frac{3H_0^2}{8\pi G} = 8.098 h^2\times 10^{-47} GeV^4$, and $H_0 = 100 \, h \,{\rm km \, s^{-1} Mpc^{-1}}$ is the Hubble parameter at the present epoch~\cite{Book-Dodelson-2020,Book-Baumann-2022}. 
In the early RD epoch where $T_{\rm end} \gg T$, the Friedmann equation provides the relation, $T_{\rm end} = \left(\frac{90 M_P^2 H_{\rm end}^2 }{\pi^2 g_* } \right)^{1/4} $. 
Upon using $\rho_{GW} \propto a^{-4}$ with $\frac{a(T_0)}{a(T)} = \left( \frac{g_s (T)}{g_s (T_0)} \right)^{1/3} \frac{T}{T_0}$, we can estimate the energy spectrum of the produced left-handed GWs at the present epoch as
\begin{equation}\label{gw-spectrum-general-relation}
\Omega_{GW,0} \simeq \frac{1}{120\pi^2} \left( \frac{90}{\pi^2} \right)^{3/4} g_*^{1/4} (t) \left( \frac{g_{s,0}}{g_s (t)} \right)^{4/3} \frac{T_g^4 (t_0)}{\rho_c^0} \left(\frac{H_{\rm end}}{M_P}\right)^{1/2} \left( \frac{\omega (t) }{T_g (t) } \right)^6   \frac{ \Omega_B}{e^{\omega (t)/T_g(t)} - 1}
\end{equation}
\begin{figure}[!t]
	\centering
	\includegraphics[height=2.8in]{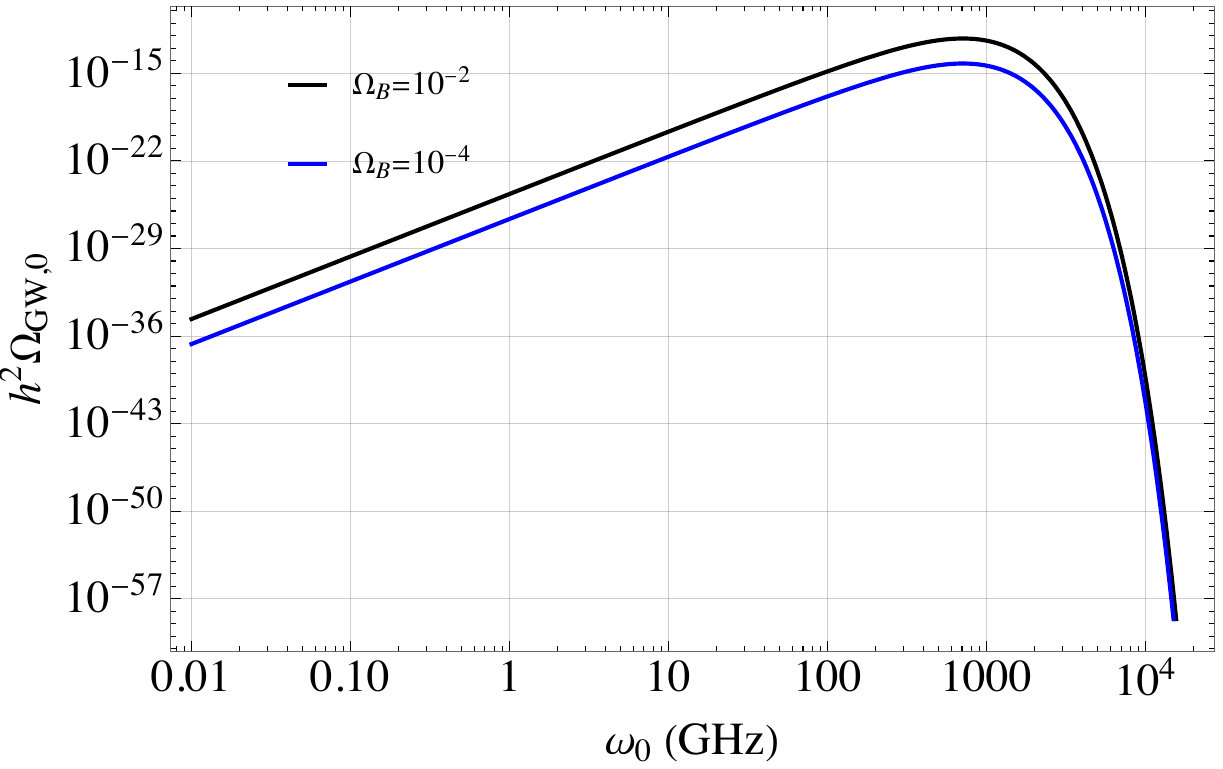} 
\caption{Energy spectrum of the produced (left-handed) gravitons for two different values of $\Omega_B$ which stays below the BBN bound, $h^2 \Omega_{GW,0} < 1.12\times10^{-6}$~\cite{2020-Aggarwal.etal-LivingRevRel}.}
\label{fig:energy-spectrum}
\end{figure}
Since both $T$ and $\omega$ are inversely proportional to the scale factor giving $\omega(t)/T_g(t) = \omega_0/T_g(t_0)$, we can write Eq.\eqref{gw-spectrum-general-relation} in terms of the comoving frequency as
\begin{align}\label{gw-spectrum-today-final}
	h^2 \Omega_{GW,0} \simeq 4.11\times 10^{-8} \left( \frac{g_* }{g_s^{32/3} } \right)^{1/4} \left(\frac{H_{\rm end}}{10^{14} \, {\rm GeV}}\right)^{1/2} \left( \frac{\omega_0 }{T_{g,0} } \right)^6    \frac{ \Omega_B }{e^{\omega_0 /T_{g,0}} - 1} 
\end{align}
where $g_*  \simeq g_s \simeq 105.25 $ and $g_{s,0} = 3.93$~\cite{2018-Saikawa.Shirai-JCAP}, and we take the Hubble parameter at the onset of RD epoch, $H_{\rm end} = 10^{14} \, {\rm GeV}$. \ref{fig:energy-spectrum} shows the energy density of the produced left-handed gravitons. 
It is evident that the energy spectrum peaks at high-frequency range, specifically $\omega_0 \sim 998 \,{\rm GHz}$, and is sensitive to the entropy relativistic degree of freedom $g_s$ at the time of conversion. Also, as expected the amplitude is larger for the stronger background magnetic field. 
Although, we have derived the energy density spectrum~\eqref{gw-spectrum-today-final} for the left-handed graviton (corresponding to case \ref{subsec:non-helical-B}), following the same assumptions and approximations one can derive the energy spectrum for both the modes of GWs by using the generic result for the conversion probability~\eqref{case2-conversion-probability}. However, the result would follow Eq.\eqref{gw-spectrum-today-final} by replacing $\Omega_B$ by $\sqrt{\delta_B} \Omega_B^+$ for left-handed mode and by $\delta_A^2 \sqrt{\delta_B} \Omega_B^+$ for right-handed mode.
\begin{figure}[t]
\centering
\includegraphics[height=3.8in,width=6in]{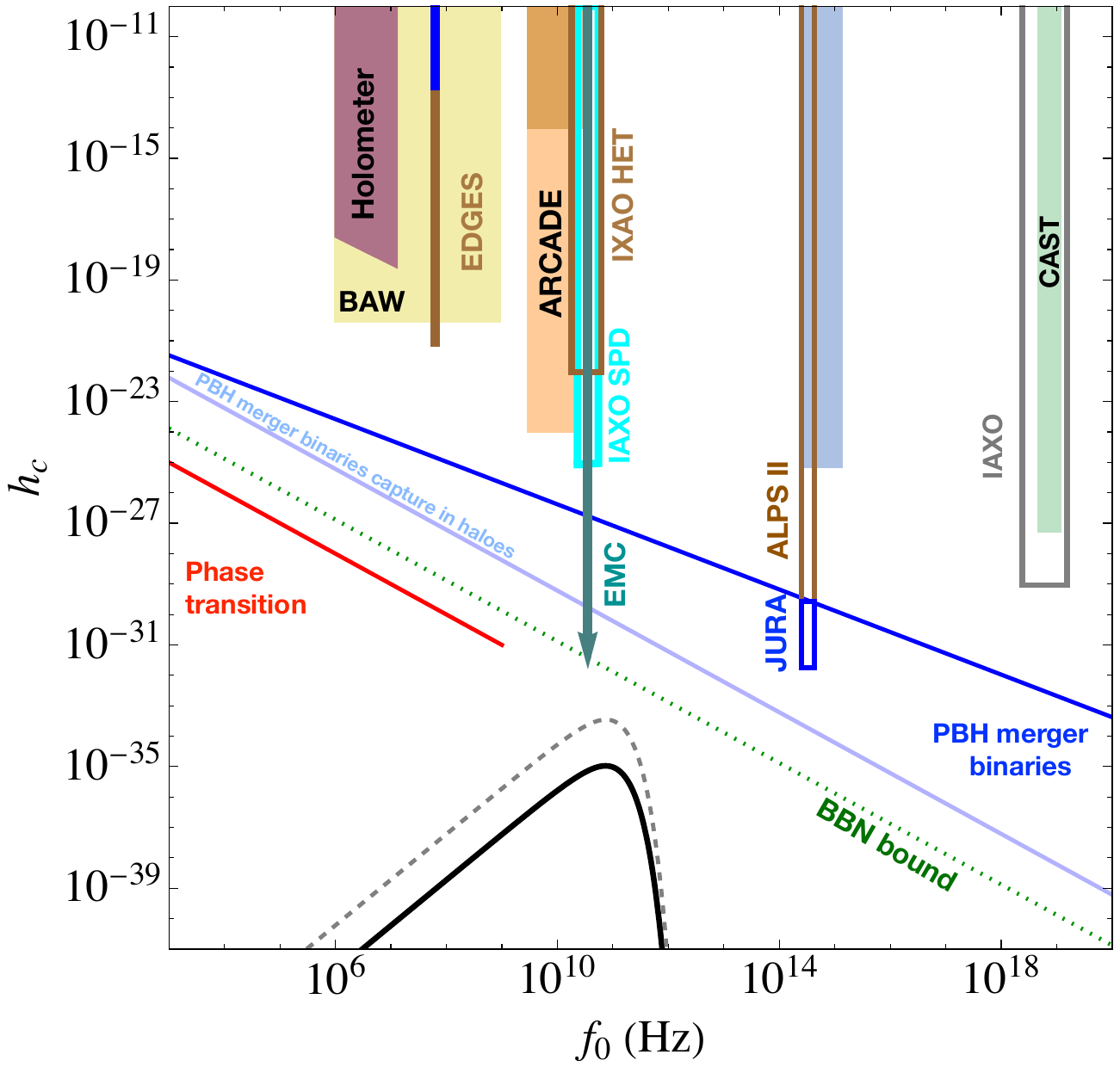} 
\caption{Comparison of the characteristic strain of produced (left-handed) GWs and other sources with sensitivity regions of ongoing/proposed detectors. The black line is for $\Omega_B = 10^{-2}, g_* = g_s = 105.25$ and the dashed grey curve is for non-standard values of relativistic degrees of freedom, $g_s=100, g_* = 10^{10}$ with $\Omega_B = 0.1$. The green dotted line indicates the BBN bound
~\cite{2020-Aggarwal.etal-LivingRevRel}. }
\label{fig:HFGW-plot}
\end{figure}
Next, we compare the prediction of our work with other sources of high-frequency GWs. To do that, we compute the characteristic GW amplitude $h_c$ (shown in \ref{fig:HFGW-plot}) for Eq.\eqref{gw-spectrum-today-final} by using the relation 
$\Omega_{GW,0} = \frac{2\pi^2}{3 H_0^2
} f^2 h_c^2$, where $f = \omega/2\pi $ is the frequency of GWs~\cite{2018-Caprini.Figueroa-CQG}. \ref{fig:HFGW-plot} also shows various detector concepts and experimental possibilities to detect high-frequency GWs, such as Bulk Acoustic Wave Devices (BAW), holometer experiment, IAXO Single Photon Detector (SPD), IAXO Heterodyne radio receiver (HET), OSQAR, CAST, ALPS II, JURA, EDGES and ARCADE, and enhanced magnetic conversion (EMC), see Ref.~\cite{2020-Aggarwal.etal-LivingRevRel} for details. To compare the prediction of GW strain from our mechanism, we also plot GWs from phase transition (red colour) and PBH merger binaries (capture in haloes) in blue (light blue) colour.
As we can see from \ref{fig:HFGW-plot}, the characteristic strain of produced left-handed GWs (black curve) has a peaked structure but with a very small amplitude to be detected in near-future missions. However, in beyond standard model (BSM) physics scenarios where $g_s=100, g_* = 10^{10}$ with $\Omega_B = 0.1$ (grey dotted curve), we can get a significantly larger strain amplitude. Consequently, the peak can be pushed towards the EMC for a given BSM scenario.
Thus, the GWs produced in such a scenario can be used as a novel tool to probe the BSM physics of the very early Universe.
\section{Estimation of the net circular polarization of GWs}\label{sec
:circular-polarization}
The properties of the chiral GWs are described by two key quantities -- the energy spectrum and net circular polarization. 
After calculating the conversion probability and the energy spectrum for the produced GW, we now compute the circular polarization parameter of the produced gravitons. 
The spectral energy density of the GWs is related to the power spectrum as $\Omega_{GW} \sim \sum_{\lambda=L,R}^{} P^{\lambda}_g$, where $\frac{2\pi^2}{k^3} P^{\lambda} = \langle h_{\textbf{k},\lambda} h_{\textbf{-k},\lambda} \rangle $~\cite{2018-Caprini.Figueroa-CQG}. 
Therefore, similar to the EM sector, we can define the parameter $\delta_g$ for the produced GW as $|h_L| = \sqrt{\delta_g} |h_R|$ or equivalently $\delta_g = \frac{\Omega_{GW}^L}{\Omega_{GW}^R}$. With these definitions, we can obtain the parameters $\Delta \chi_g$ and $\Delta \chi_{\gamma} $, which describe the net-circular polarization of the produced GWs and propagating EM waves as~\cite{2011-Sorbo-JCAP,2010-Gluscevic.Kamionkowski-PRD}
\begin{align}\label{handedness-parameter-def}
\Delta \chi_g \equiv \frac{P^R_g - P^L_g}{P^R_g + P^L_g} = \frac{1-\delta_g}{1+\delta_g} \quad \text{and} \quad \Delta \chi_{\gamma} \equiv \frac{P^R_{\gamma} - P^L_{\gamma} }{P^R_{\gamma} + P^L_{\gamma} } =  \frac{1-\delta_A}{1+\delta_A}
\end{align}
where $P^{L,R}$ refer to the power spectrum of the corresponding mode. Note that $\Delta \chi_{g,\gamma} = + 1 (-1)$ refers to the maximally helical right-handed (left-handed) wave. 

\begin{figure}[t]
\centering
\includegraphics[height=3in,width=3.8in]{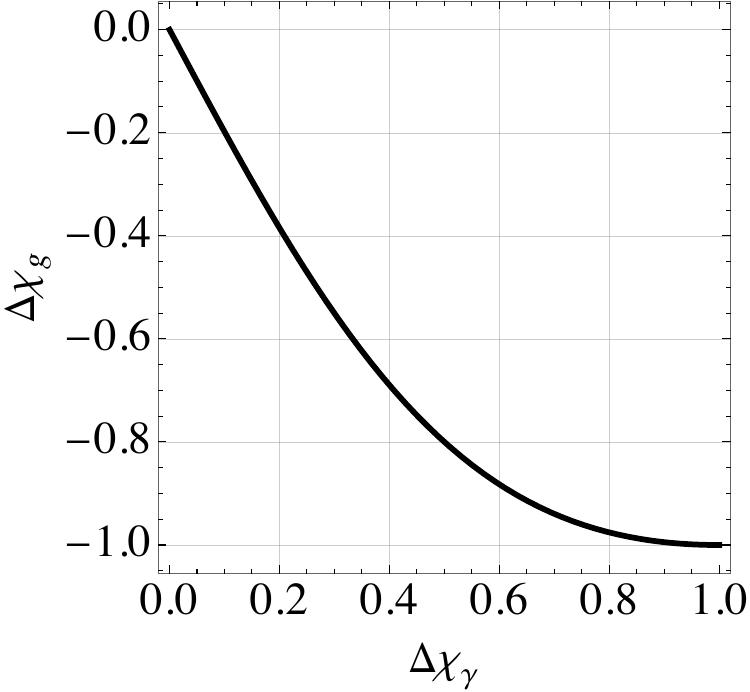} 
\caption{Behaviour of the circular polarization parameter $\Delta \chi_g$  of produced chiral GWs as a function of $\Delta \chi_{\gamma}$ of EM waves.}
\label{fig:circular-polarization}
\end{figure}

To calculate the relation between $\Delta \chi_g$ and $\Delta \chi_{\gamma}$, 
let us now consider Eq.\eqref{case2-conversion-probability}, which gives the conversion probability for the generic case where both propagating EM wave and background magnetic field could be helical. Following the calculations and assumptions used to obtain Eq.\eqref{gw-spectrum-general-relation}, we can derive the relation between the energy density carried by both the helical modes of the produced GWs as
\begin{equation}\label{delta-A-g-relation}
		\Omega_{GW,0}^R =  \delta_A^2 \,\,  \Omega_{GW,0}^L \quad \implies \quad \delta_g = \frac{1}{\delta_A^2}
\end{equation}
where in obtaining the last relation we used definitions of the corresponding parameters. Now, using Eq.\eqref{handedness-parameter-def} and Eq.\eqref{delta-A-g-relation}, we obtain 
\begin{equation}\label{chirality-parameter-relation-main}
	\Delta \chi_g =  -\frac{2 \Delta \chi_{\gamma}}{1+\Delta \chi_{\gamma}^2} ~~.
\end{equation}
This expression is one of the main results of this work, which gives the relation between the chirality of the produced GWs in terms of chirality of the propagating EM waves, and shown in \ref{fig:circular-polarization}.  
As we can see, Eq.~\eqref{delta-A-g-relation} for the maximally helical propagating EM waves with $\Delta \chi_{\gamma} \simeq +1$ (i.e., $|A_+| \gg |A_-|$), the produced GWs are maximally helical left-handed with $ \Delta \chi_g \simeq -1$, which is consistent with our previous case~(\ref{subsec:non-helical-B}). In Eq.\eqref{chirality-parameter-relation-main}, we can see that the chiral nature of the background magnetic field does not affect the chirality of produced GWs, which shows that the background magnetic field (the catalyst) democratically affects both the modes.

Before concluding this section, we would like to differentiate the chiral GWs produced in this work from other mechanisms where parity violation is either in the gravitational sector producing the chiral GWs (primary)~\cite{2009-Alexander.Yunes-PhyRept} or in the matter sector (sourcing the second order chiral GWs)~\cite{2011-Sorbo-JCAP,2018-Yoshida.Soda-IJMPD}. While this resonant mechanism necessarily requires a background magnetic field, other mechanisms do not. Furthermore, in our work, the model-dependent predictions from the EM sector can be studied by following the same analysis with a suitable choice of the coupling function $f(\phi)$. While our mechanism relates the chirality of GW with the chirality of the EM waves in a model-independent way as in Eq.~\eqref{chirality-parameter-relation-main}, other scenarios usually exhibit a model-dependent definition~\cite{2011-Sorbo-JCAP}.
In contrast to other mechanisms, the chirality production in the GW sector from the EM sector is oscillatory in nature due to the photon-graviton conversion (and vice-versa). Since the probability of conversion is small even for a larger background magnetic field (see \ref{fig:conv-probability}), the conversion of the produced chiral gravitons to the chiral photons, would be highly suppressed. Note that this chirality oscillation is different than often referred to in the literature as the chirality oscillation between the amplitudes of the left- and right-handed GWs modes converting into each other and oscillating in their propagations (produced from the parametric resonance amplification of GWs)~\cite{2017-Cai.etal-JHEP}. In these mechanisms, the chirality parameter of GW has oscillatory features with a peak determined by the model's parameter space.
In Refs.~\cite{2020-RoperPol.Kahniashvili.etal-PRD,2021-Kahniashvili.Brandenburg.etal-PhyRevRes}  authors performed a numerical analysis and showed that post-inflationary physics associated with magnetohydrodynamic turbulence, in the presence of helical initial magnetic fields, can also give rise to net circular polarization of an SGWB potentially detectable with LISA. However, this is completely different from the photon-graviton conversion mechanism discussed in this work.

\section{Conclusions and discussions}

GWs are one of the most powerful tools for probing the physics of the early universe.
While the amplitude of GWs reflects the efficiency of their generation, the presence of non-vanishing circular polarization indicates the parity-violating nature of the mechanisms behind their production.
In this work, we explored how chirality is transferred from the EM sector to the GW sector through the photon-graviton conversion during the RD era.
In our model-independent setup, due to the parity-violating nature of the system which includes propagating EM waves and background magnetic field, the left- and right-handed modes of the produced GWs have diﬀerent amplitudes. Therefore, the produced GWs are chiral in nature with a net circular polarization, which leads to a manifestation of parity violation in the GW sector transferred from the EM sector, with the conversion mechanism being most effective in stronger magnetic fields.
Further, we derived an expression that relates the net circular polarization of the produced GWs to the propagating EM waves. Our results indicate that the chirality parameter is insensitive to the chirality of the background magnetic field and depends solely on the chirality of the propagating EM waves. We have also highlighted the differentiating features of the chirality relation in this work from other mechanisms.
Moreover, the energy spectrum of the generated chiral GWs has a peak frequency $\sim 100$ GHz, which is also sensitive to the relativistic degrees of freedom at the time of conversion. The power spectrum of the background magnetic fields characterised by a certain coherence length would introduce a length scale into the photon-graviton systems, i.e., only photons with a wavelength smaller than the coherence length of the magnetic field would be converted to gravitons, which might affect the efficiency of the conversion and consequently, shifts the peak frequency of the produced chiral GWs.
To assess the significance of the energy contribution of these produced gravitons to the total stochastic GW background, we compared the characteristic strain of the generated GWs with other high-frequency GW sources and contrasted our predictions with the sensitivities of current and future missions aimed at detecting high-frequency GWs.

Detection of high-frequency GWs would open a new observational window, complementing the insights gained from the EM and low-frequency GW observations.
Detecting these GWs has been one of the key objectives of various future GW missions. 
However, their detection presents unique challenges due to their weak interactions and extremely small amplitudes. 
Advanced techniques and highly sensitive detectors based on superconducting circuits and interferometric devices are being developed to capture these extremely weak signals. 
Although many different mechanisms can produce GWs in the high-frequency regime, the scenario explored in this work is a unique model-independent mechanism that can produce high-frequency GWs along with a net non-zero chirality. 
Since the peak frequency of produced GWs depends on the relativistic degrees of freedom, this mechanism can potentially probe the non-standard cosmic history of the Universe. 
Future missions and advancements in technology are expected to enhance our ability to detect such high-frequency GWs, clearly revolutionizing our understanding of high-energy astrophysics and early universe cosmology. Although the mechanism explored here takes place in the early universe, our analysis can be used at any epoch with suitable approximations relevant to a given physical situation, such as in neutron stars, magnetars and white dwarfs. We are currently exploring these interesting directions.
\section*{Acknowledgements} \label{sec:acknowledgements}
AK and RKJ thank Valerie Domcke for pointing out an important issue.
AK and RKJ acknowledge financial support from the Indo-French Centre for the Promotion of Advanced 
Research (CEFIPRA) for support of the proposal 6704-4 under the Collaborative Scientific 
Research Programme.
RKJ also acknowledges support from the IISc Research Awards 2024 and SERB, Department of Science and Technology, GoI through the MATRICS grant~MTR/2022/000821. 
AK and RKJ also thank Basundhara Ghosh, Brijesh Kanodia, Jishnu Sai P, Rathul N Raveendran, Subhadip Bouri and Yashi Tiwari for useful discussions and insightful comments on the work.

\input{References.bbl}
\bibliographystyle{apsrev4-1}
\end{document}

%% file: References.bbl
%